\documentclass[aps,prl,twocolumn,showpacs,groupedaddress]{revtex4}
\usepackage{graphicx}

\begin{document}

% \draft command makes pacs numbers print
%\draft

\title{Inhibiting Three-Body Recombination in Atomic Bose-Einstein Condensates}
\author{Chris P. Search$^1$, Weiping Zhang$^{1,2}$, and Pierre Meystre$^1$}
\affiliation{$^1$Optical Sciences Center, University of Arizona,
Tucson, AZ 85721 \\$^2$Department of Physics,
Tsinghua University, Beijing, 100084, P. R. China}

\begin{abstract}
We discuss the possibility of inhibiting three-body recombination in atomic Bose-Einstein 
condensates via the application of resonant $2\pi$ laser pulses. These pulses 
result in the periodic change in the phase of the molecular state by $\pi$, 
which leads to destructive interference between the decay amplitudes following 
successive pulses. We show that the decay rate can be reduced by several orders
of magnitude under realistic conditions. 
\end{abstract}

\maketitle

A fundamental limit to the lifetime of trapped atomic Bose-Einstein condensates 
(BECs) is three-body recombination, in which two atoms form a molecular dimer and the third atom 
carries away the excess energy and momentum \cite{kagan,fedichev,esry,jack,borca}. Unlike other loss 
mechanisms such as spin relaxation, collisions with background thermal atoms, or spontaneous light
scattering, three-body recombination is an intrinsic loss mechanism that cannot be eliminated by simply engineering 
a better trapping environment.

The decay rate due to three-body recombination is proportional to the square of the atomic density,
\begin{equation}
\frac{d n }{dt}=-3K_3 n ^3 /6 \label{K3}
\end{equation}
where $n$ is the atomic density, $K_3\sim \hbar 
a^4/m$, and $a$ is the elastic atomic $s$-wave scattering length for atoms of mass $m$ 
\cite{fedichev,esry,burt,stamper-kurn,stenger,roberts,weber}. (We note that Eq. (\ref{K3})
contains an additional factor of $1/6$ if all atoms occupy the same quantum state \cite{kagan}).  
At high densities and large scattering lengths, three-body recombination represents the primary 
limit on the lifetime of condensates. 

The strong dependence of three-body recombination on $a$ and 
the density imposes severe restrictions on experiments done in the vicinity of a Feshbach resonance 
\cite{stenger,roberts,weber}, where $a$ diverges, as well as for tightly confined samples. The latter 
situation is relevant in particular to the new field of integrated atom optics \cite{hansel}. 
Thus it is important to pursue possible techniques for inhibiting three-body recombination in condensates.

The control of decay and decoherence mechanisms in quantum systems is also important in other areas of physics 
\cite{search,agarwal2}. Much of this work is motivated by the requirements for quantum computers as well as an interest in fundamental 
issues in quantum mechanics, such as the Quantum Zeno effect \cite{misra}. An important result in this context is the 
demonstration that time-dependent external fields that modulate either the energies of the system under consideration 
or its coupling to the external reservoir into which it decays can reduce its decay rate. A key requirement of such 
schemes is that the external field be modulated on a time scale shorter than the reservoir correlation time.

For positive scattering lengths much larger than the range of the interatomic
potential, three-body recombination is dominated 
by the formation of molecules in a weakly bound state with binding energy $\epsilon_b=\hbar/m a^2$ \cite{fedichev,esry}.
In this paper we discuss the use of a regular sequence of 
resonant $2\pi$ laser pulses applied to that state to inhibit three-body decay. 
The effect of each pulse is to change its phase by $\pi$. This leads to a 
destructive interference in the probability amplitudes between successive 
pulses for the formation of a molecule. A similar idea was explored by Agarwal {\it et al.} for inhibiting 
spontaneous emission from a two-level atom \cite{agarwal2}. However, it is virtually impossible to inhibit 
spontaneous emission in free space because of the nearly instantaneous 
correlation time of the vacuum. The situation is more favorable in the present case: The role of the reservoir 
is now played by the molecular states formed by three-body recombination, and its correlation time is roughly on the order of 
the inverse of the binding energy, $\epsilon_b^{-1}\sim 10^{-6}s$ for $a\sim 1000a_0$ and $m\sim 50$ a.m.u. Therefore a 
sequence of pulses separated by an interval of $T\ll \epsilon_b^{-1}$ should be 
effective at decreasing the rate of molecule formation.

We proceed by deriving a coarse-grained master equation for the evolution of a BEC 
in the presence of a sequence of impulsive laser pulses. The pulses are assumed to be off-resonant 
with respect to the condensate atoms, but resonant with respect to the molecules 
formed by three-body recombination. The total Hamiltonian for the atom-molecule 
system is $H=H_a+H_m+H_3$ where
\begin{widetext}
\begin{eqnarray}
H_a &=&\int d^{3}x \left\{ \hat{\psi}^{\dagger}\left(-\frac{\hbar^2 \nabla^2}{2m}+V({\bf 
r})-\mu \right)\hat{\psi} +\frac{g}{2}\hat{\psi}^{\dagger}\hat{\psi}^{\dagger}\hat{\psi}\hat{\psi} \right\}, 
\,\,\,\,\,\,\,\,\,\,\,\,\,\,  H_3=\hbar\kappa\int d^{3}x \hat{\phi}_g^{\dagger}\hat{\psi}^{\dagger}\hat{\psi}^3 +h.c.\nonumber \\
H_m&=&\int d^{3}x \left\{ 
\hat{\phi}_g^{\dagger}\left(-\frac{\hbar^2\nabla^2}{4m}-\hbar\epsilon_b\right)\hat{\phi}_g + 
\hat{\phi}_e^{\dagger}\left(-\frac{\hbar^2\nabla^2}{4m}+\hbar(\omega_{eg}-\epsilon_e-\omega_L) 
\right)\hat{\phi}_e +
\frac{\hbar}{2}\Omega_m(t)\hat{\phi}_e^{\dagger}\hat{\phi}_g +h.c. \right\} \nonumber
\end{eqnarray}
\end{widetext}
Here $H_a$ and $H_m$ are the atomic and molecular Hamiltonians, respectively, and $H_3$ 
accounts for three-body recombination \cite{kagan}. The field operators $\hat{\psi}({\bf r})$ and $\hat{\phi}_g({\bf r})$ 
describe the annihilation of atoms and of molecules with 
binding energy $\epsilon_b$, respectively. In $H_a$, $g=4\pi\hbar^2 a/m$ while $V({\bf r})$ 
is an external trapping potential and $\mu$ is the chemical potential.  

We assume that the laser field, with Rabi frequency $\Omega_m(t)$ and frequency $\omega_L$, is resonant 
with a single vibrational state in the molecular potential of the electronically excited molecule. This state
is denoted by the annihilation operator $\hat{\phi}_e({\bf r})$. Its binding energy relative to the electronic 
energy of the corresponding free atoms is $\epsilon_e$, so that the resonance condition is 
$\delta=(\omega_{eg}-\epsilon_e)+\epsilon_b-\omega_L=0$. We further assume that the laser is 
off-resonant with respect to the excited state transition for the corresponding free atoms, 
$\Delta_a=\omega_{eg}-\omega_L\gg \Omega_m/\sqrt{2},\gamma_a$, where $\Omega_m/\sqrt{2}$ is 
the atomic Rabi frequency \cite{heinzen} and $\gamma_a$ the linewidth. 
For $\delta=0$ this gives $\Delta_a=\epsilon_e-\epsilon_b$. Since $\epsilon_b$ corresponds 
to a weakly bound state near the dissociation limit, $\epsilon_e$ would have to correspond 
to a low lying vibrational state in the molecular potential for the excited atoms with a binding energy of the order
$\sim 1-10$GHz.

For the problem at hand, $\Omega_m(t)=\Omega_0f(t)$ where $f(t)$ is a train of square pulses of unit amplitude 
with duration $\tau_p$ and separation $T$ such that $\Omega_0\tau_p=2\pi$. 
If $\tau_p$ is short compared to the characteristic time of the center-of-mass molecular dynamics, the molecules 
undergo a complete Rabi oscillation for each pulse, leaving their excited state unchanged while their ground state acquires a $\pi$ phase shift.
We can then eliminate the excited molecular state from $H_m$ and make for the ground state field operators the substitution
\[
\hat{\phi}_g({\bf r},t)\rightarrow(-1)^{\left[\frac{t}{T+\tau_p}\right]}\hat{\phi}_g({\bf 
r},t)
\]
where $[...]$ denotes the integer part of the term in brackets.

The Hamiltonian $H_3$ describes the formation of dimers from three colliding atoms in which the third atom
carries away the excess kinetic energy and momentum released by the molecule formation. The binding energy of the molecule 
is converted into molecular and atomic kinetic energies, $\epsilon_b=3\hbar K^2/4m$ where $+\hbar {\bf K}$ and  $-\hbar {\bf K}$ are the 
atomic and molecular momenta. For $\hbar K^2/4m\gg \mu$, we can 
decompose the atomic field operator as $\hat{\psi}({\bf r})=\hat{\psi}_T({\bf r})+\hat{\psi}_F({\bf 
r})$ where $\hat{\psi}_T$ represent the trapped condensate atoms while $\hat{\psi}_F$ are atoms with kinetic energy $2\epsilon_b/3$. 
By making what amounts to the rotating wave approximation and keeping only 
resonant terms, $H_3$ reduces to \cite{jack},
\[
H_3=\hbar\kappa\int d^{3}x \hat{\phi}_g^{\dagger} \hat{\psi}_F^{\dagger} \hat{\psi}_{T}^{3} +h.c.
\]

We note that the molecules and recoiling atoms escape from the trap for $\epsilon_b/3 > V_0$ 
where $V_0$ is the trap depth. For large scattering lengths with $\epsilon_b\sim 10^{6}s^{-1}$, 
atoms and molecules are lost for $V_0/k_B\lesssim 10\mu K$. 
Under these conditions, it is convenient to adopt a plane wave basis 
for the atoms and molecules that are lost from the trap,
$\hat{\phi}_g({\bf r},t)=(1/\sqrt{V})\sum_{\bf k}\hat{a}_{\bf k}(t)\exp[i{\bf k}\cdot 
{\bf r}]$ and $\hat{\psi}_F({\bf r},t)=(1/\sqrt{V})\sum_{\bf k}\hat{c}_{\bf k}(t)\exp[i{\bf k}\cdot {\bf r}]$,
where $V$ is a quantization volume. For the condensate we use a zero-temperature single-mode approximation, 
$\psi_T({\bf r},t)=u_0({\bf r})\hat{b}(t)$ and $u_0({\bf r})$ is a Hartree wave function for the condensate 
ground state with eigenvalue $\mu$. After transforming to the interaction representation, the interaction Hamiltonian becomes
\begin{equation}
H_3(t)=\sum_{ {\bf k}_1, {\bf k}_2}\hbar U({\bf k}_1+{\bf 
k}_2) (-1)^{[t/T+\tau_p]}e^{i\delta\omega_{12}t} \hat{a}^{\dagger}_{{\bf k}_1}\hat{c}^{\dagger}_{{\bf 
k}_2}\hat{b}^3 +h.c \label{H3}
\end{equation}
where $\delta\omega_{12}=\hbar {\bf k}_1^2/(4m)+\hbar {\bf k}_2^2/(2m)-\epsilon_b$
and $U({\bf k})=\kappa\int d^{3}x e^{-i{\bf k}\cdot{\bf r}}u_0^3({\bf r})/V$.

Starting from Eq. (\ref{H3}) we can derive a master equation for the density operator of the
condensate atoms, $\rho(t)$. In this approach the molecules and free atoms are interpreted as a reservoir 
coupled to the condensate by $H_3(t)$. It is important to note that $H_3(t)$ is not a continuous function 
of time since the pulses result in a discontinuous sign change after each interval $T+\tau_p$. 
Hence one must proceed with caution when deriving the master equation  
since differentiation and integration are no longer inverses of each other. We use the form of the master 
equation derived e.g. in Ref. \cite{meystre} for the coarse-grained derivative $\dot{\rho}(t)=(\rho(t+\tau)-\rho(t))/\tau$ 
of the system density operator, $\tau$ being a time long compared to the correlation time of the reservoir $\tau_c$ 
but short compared to times over which the condensate evolves, $\tau_c\ll \tau \ll 1/\gamma n^2$. Taking the 
reservoir density operator to be in the vacuum state and assuming that $\tau= NT$ where $N$ is the number of 
pulses and $\tau_p\ll T$, we obtain
\begin{eqnarray}
\dot{\rho}(t)&=&\frac{1}{i}\left[\nu\hat{b}^{\dagger 3}\hat{b}^3, \rho(t)\right] \nonumber \\
&-&\frac{\gamma}{2}\left(\hat{b}^{\dagger 
3}\hat{b}^3\rho(t) +\rho(t)\hat{b}^{\dagger 3}\hat{b}^3 
-2\hat{b}^3\rho(t)\hat{b}^{\dagger 3} \right).
\end{eqnarray}
Here $\nu\equiv \nu(T,N)$ is a ``Lamb shift'' of the condensate atoms. We do not reproduce its lengthy expression, concentrating 
instead on the decay rate of the condensate, $\gamma\equiv\gamma(T,N)$ due to molecule formation,
\begin{eqnarray}
\gamma(T,N)&=&\sum_{{\bf k}_1, {\bf k}_2} \left| U({\bf k}_1+{\bf k}_2) \right|^2 
\tan^2(\delta\omega_{12}T/2) \nonumber \\
&\times& \left ( \frac{\sin^2\left(\delta\omega_{12}NT/2 +N\pi/2 \right)}{NT \delta\omega_{12}^2/4}\right).
\end{eqnarray}

The $\tan^2 \left(\delta\omega_{12}T/2 \right)$ term in $\gamma(T,N)$ describes the effect of 
the pulses on the decay of the condensate. We note that $\gamma(T,N)NT$ agrees with the transition probability 
calculated directly from $H_3(t)$ using first-order time-dependent perturbation 
theory \cite{agarwal2}.

For a Markovian reservoir, $\tau_c\to 0$, one can let 
$\tau=NT\to\infty$ \cite{meystre}. At the same time we note that for an even 
number of pulses no net phase (modulo $2\pi$) is acquired by the molecules. Under these conditions the decay rate is
\begin{equation}
\lim_{\tau\to\infty}\gamma=2\pi \sum_{{\bf k}_1, {\bf k}_2} \left| U({\bf k}_1+{\bf k}_2)\right|^2
\tan^2 \left(\delta\omega_{12}T/2 \right) \delta( \delta\omega_{12}). 
\label{markov_decay}
\end{equation}
For $\tan^2 \left(\delta\omega_{12}T/2 \right)\equiv 1$, one then recovers the 
standard result for the decay rate in the Markov limit, as expected. 

It is clear from Eq. (\ref{markov_decay}) that $\lim_{\tau\to\infty}\gamma=0$. 
This is due to the fact that for $\delta\omega_{12}=0$, the phases of the 
molecules and free atoms do not change in the interval between pulses, 
$j(T+\tau_p)<\tau < (j+1)T$, where $j$ is an integer. However at the end of each interval, 
$\tau=j(T+\tau_p)$, the phase of the molecules changes by $\pi$. As a result, there is complete destructive 
interference between the transition amplitudes for neighboring intervals. 

For a finite reservoir correlation time $\tau_c$, however, it is no longer possible to (formally) let 
$\tau=NT\to\infty$ and the interferences cease to be fully destructive, resulting in a non-zero condensate decay rate. 
To reduce $\gamma$ below its unperturbed value requires $\tan^2 \left(\delta\omega_{12}T/2 \right)\ll 1$ 
but since $\tau_c$ is approximately given by the reciprocal of the reservoir bandwidth, $\epsilon_c$,
one must choose $T\ll \pi\tau_c$, that is, the laser pulses must be separated by an interval shorter than the 
correlation time of the molecular field. The bandwidth of the molecular reservoir is determined by the 
short-wavelength cutoff in $H_3$, which is on the order of the range of the 
interatomic potential, $R_0\lesssim a$ \cite{kagan,bedaque}, This gives a bandwidth of 
$\epsilon_c\sim\hbar/mR_0^2\gtrsim\epsilon_b$.

The de Broglie wavelengths of the molecules and energetically free atoms are of the order $a$.
For a trapped condensate of size $\ell$, one typically has $\ell \gg a$. 
In this case, we may treat the gas as being locally homogeneous with respect to the fast moving molecules
and atoms formed by recombination. For a uniform BEC, $\gamma$ then becomes
\begin{equation}
\gamma=\gamma_0F(T\epsilon_b,N)
\end{equation}
where $\gamma_0=(1/2\pi)(\kappa/V)^2(4m/3\hbar)^{3/2}\sqrt{\epsilon_b}$ is the decay 
rate in the absence of any pulses in the $\tau\to\infty$ Markov limit. Note that $\gamma$ is related to $K_3$ by 
$K_3=36\gamma$ \cite{jack}. The dimensionless factor reflecting the effect of the $2\pi$  
pulses is
\begin{equation}
F(t,N)=\int^{\alpha}_{-1}dx \sqrt{1+x} \frac{\tan^2 
(xt/2)\sin^2((xt+\pi)N/2)}{2\pi Nt(x/2)^2}. \label{FT}
\end{equation}
Here $\alpha=(\epsilon_c-\epsilon_b)/\epsilon_b$ is a dimensionless high energy cutoff.
By setting $\tan^2(xt/2)=1$ in Eq. (\ref{FT}) we obtain the {\em finite} time decay rate
in the absence of any pulses, which we denote as $F_{NP}(t,N)$. $F_{NP}$ rapidly approaches its 
limiting value of $1$ for $t\sim 1$ for all values of $N$ and $\alpha$ considered 
below. This is illustrated in Fig. 1 for $N=20$ and $\alpha=1.5$

\begin{figure}
\includegraphics*[width=8cm,height=4cm]{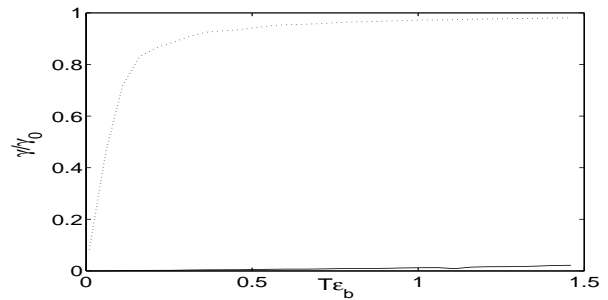}
\caption{Normalized decay rate for $N=20$ and $\alpha=1.5$ with pulses (solid line) and without pulses (dotted line)
as a function of $T\epsilon_b$ }
\label{fig1}
\end{figure}

The effective number of pulses $N=[\tau/T]$ that contribute to the reduction of $\gamma$
is determined by the time over which the phases of the molecular states can evolve coherently. 
Any event that leads either to the decay of the molecular state or to a randomization of its phase
will negate the accumulated effect of the pulses. Since the molecules formed by three-body recombination
are in a very weakly bound state, they can decay to more deeply bound 
vibrational states via inelastic collisions with atoms \cite{yurovsky, timmermans,soldan,balakrishnan}.
The decay rate for the molecular state is then given by $\kappa n=\beta n\bar{v}\sigma\approx 4\pi\beta n\hbar a/\sqrt{3}m$. 
Here, $n\bar{v}\sigma$ is the elastic collision rate between atoms and molecules with velocity 
$\bar{v}=\hbar/\sqrt{3}ma$ and we assume that the cross section can be approximated by the atom-atom 
elastic cross section, $\sigma\approx 4\pi a^2$. $\beta$ is then the ratio of the inelastic 
to elastic cross sections for transitions to deeply bound vibrational states of 
the molecules.

We can therefore take $\tau \approx 1/\kappa n$ in order to determine the effective number of 
pulses. In this case the condition on the time scales involved in the 
derivation of the master equation, $1/\gamma n^2 \geq 1/\gamma_0n^2 \gg \tau \gg \tau_c $, 
can be reexpressed as $na^3\ll 4\pi\beta/\sqrt{3}, \sqrt{3}/(4\pi\beta) (a/R_0)^2$ where $na^3$ 
is the dilute gas parameter and is typically $\ll 1$. Refs. \cite{yurovsky,timmermans,soldan,balakrishnan} give 
empirical values for $\kappa$ of $10^{-9}-10^{-11}$ cm$^{3}$/s in the vicinity of a Feshbach resonance. 
Using the values $a\sim 1000a_0$ and $m\sim$ 50a.m.u. gives a range for $\beta$ of $0.01-1$. 
With these values of $\beta$ one can reduce the above condition on the time scales for $a>R_0$ to simply $na^{3}\ll 0.1$ 
so that the condensate must simply be dilute.

These considerations imply that for a condensate density of $n=10^{15}$/cm$^3$ and $\kappa\approx 10^{-10}$cm$^3$/s, one can have on the 
order of $10$ pulses within the lifetime of a molecule for a pulse period of $T=10^{-6}$s.
Fig. 2 shows a plot of $\gamma/\gamma_0$ as a function of $T$ for $N=10,11,19,20$. It is clear that the
more pulses that can be applied, the greater the reduction in the decay rate. In general for $NT\epsilon_b \ll 1$ and finite,
an even number of pulses, $2j$, produces a lower decay rate than $2j\pm 1$ while for larger $T\epsilon_b$ the decay rate can be lower
for an odd number of pulses. However, the difference between an even and odd number of pulses decreases 
as $N$ gets larger. We note that $F(t,N)$ is sensitive to the precise value of the cutoff 
\cite{agarwal2}, as illustrated in Fig. 3. This is a common feature in effective low-energy field theories \cite{bedaque}. 
Despite the dependence on $\alpha$, one has $F(T\epsilon_b,N)\ll 1$ provided $T\epsilon_b,T\epsilon_b\alpha \ll \pi$. 
Fig. 3 shows that when $T\epsilon_b\alpha \approx \pi$, the decay rate starts to increase rapidly. 
From these results, we conclude that the pulse train is able to reduce the decay rate to only a few percent of 
its ``bare'' value.  

\begin{figure}
\includegraphics*[width=8cm,height=4cm]{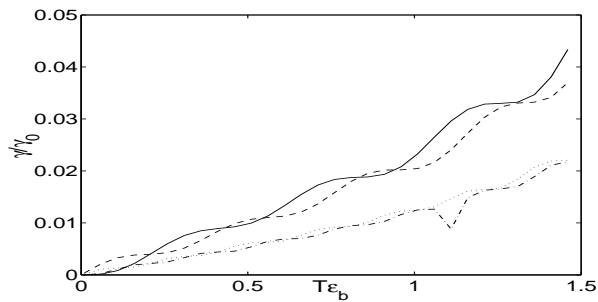}
\caption{Normalized decay rate as a function of the dimensionless pulse period, $T\epsilon_b$ for
$N=10$ (solid line), $N=11$ (dashed line), $N=19$ (dotted lined), and $N=20$ (dashed dot) for a cutoff of
$\alpha=1.5$. }
\label{fig2}
\end{figure}

\begin{figure}
\includegraphics*[width=8cm,height=4cm]{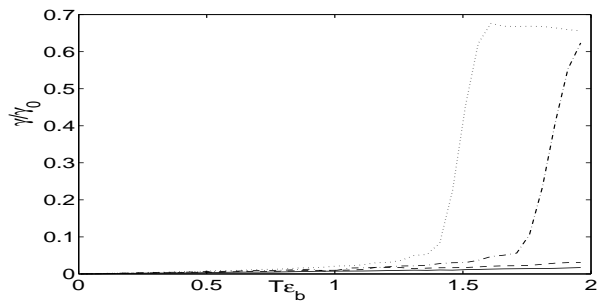}
\caption{Normalized decay rate as a function of the dimensionless pulse period, $T\epsilon_b$ for
$N=20$ and a cutoff of $\alpha=0.9$ (solid line), $\alpha=1.3$ (dashed line), $\alpha=1.7$ (dashed dot), and $\alpha=2.1$ 
(dotted lined).}
\label{fig3}
\end{figure}

Finally we discuss the effect of the laser pulses on the atoms. The atoms 
experience a periodic AC Stark shift, $\hbar|\Omega_a(t)|^2/4\Delta_a$.
It leads to a renormalization of the binding energy relative to the 
dissociation limit and can be neglected without loss of generality.
Of greater concern are the atom losses due to Rayleigh 
scattering of laser photons. They should be less than the losses due to three-body 
recombination, otherwise we have simply replaced one loss mechanism with 
another. For $t\gg T+\tau_p$, the time-averaged Rayleigh scattering loss rate is
$\bar{\Gamma}=(\gamma_a|\Omega_0|^2/8\Delta_a^2)(\tau_p/T)
\approx (0.785/T)(|\Omega_0|/|\Delta_a|)(\gamma_a/|\Delta_a|)$
where we have again assumed that $T\gg\tau_p=1/2\pi\Omega_0$. If we take $\gamma_a\sim10^7s^{-1}$, 
a detuning of $\Delta_a\sim 10^{10}s^{-1}$, and $10\tau_p=T\sim 10^{-6}s$, we 
obtain a lifetime of $\bar{\Gamma}^{-1}\sim 10$s. Note, however, that $\bar{\Gamma}$ is a single-particle loss 
rate and is unaffected by changes in the density $n$ and $a$.

In conclusion, we have discussed a technique for inhibiting three-body recombination in BEC via a sequence of 
resonant $2\pi$ laser pulses. We have shown that the three-body decay rate can be reduced to only a few 
percent of its value in the absence of the pulses. This method should be useful for extending the 
lifetime of condensates in the high density regime and near Feshbach resonances.

This work is supported in part by the US Office of Naval
Research, by the NSF, by the US Army Research Office, by NASA, 
and by the Joint Services Optics Program.

\end{document}